\documentclass[%
 reprint,
superscriptaddress,
 amsmath,amssymb,
 aps,
]{revtex4-2}

\usepackage{graphicx}
\usepackage{dcolumn}
\usepackage{bm}
\usepackage{color}
\newcommand{\ket}[1]{|#1\rangle}
\newcommand{\bra}[1]{\langle #1 |}
\newcommand{\cra}[1]{\hat{#1}^\dagger}
\newcommand{\an}[1]{\hat{#1}}

\newcommand{\iner}[2]{\langle #1 | #2 \rangle}
\newcommand{\mat}[3]{\langle #1 |#2| #3 \rangle}

\newcommand{\av}[1]{\langle #1 \rangle}

\newcommand{\m}[1]{\mathcal{#1}}
\begin{document}

\preprint{APS/123-QED}

\title{Phaseless auxiliary-field quantum Monte Carlo in the Hartree-Fock-Bogoliubov manifold}

\author{Zhaozhan Zhang}
\email{zhangchzh6@mail.sysu.edu.cn}
\affiliation{Sino-French Institute of Nuclear Engineering and Technology, Sun Yat-sen University, Zhuhai, 519082, China}
\affiliation{Department of Physics, Graduate School of Science, The University of Tokyo, Tokyo, 113-0033, Japan}
\begin{abstract}
We formulate and implement the phaseless auxiliary-field quantum Monte Carlo (AFQMC) method in the Hartree-Fock-Bogoliubov (HFB) manifold for systems with strong pairing correlations. Our formulation introduces a more flexible representation of HFB walkers with an explicit analytical expression for the normalization factor derived from the representation matrix with respect to the reference vacuum. We also extend our previously developed stochastic gauge formalism from the Slater determinant manifold to the HFB framework, providing additional flexibility for controlling stochastic fluctuations. Benchmark calculations for the Richardson model demonstrate the accuracy and numerical stability of the method.
\end{abstract}

\maketitle
\section{Introduction}

Auxiliary-field quantum Monte Carlo (AFQMC) has emerged as one of the most powerful and versatile numerical methods for investigating strongly correlated fermionic many-body systems. By reformulating the imaginary-time projection onto the ground state as a path integral over auxiliary fields, evaluated stochastically through Monte Carlo sampling, AFQMC performs a random walk in the manifold of independent-fermion wave functions, most commonly Slater determinants. This formulation circumvents the exponential scaling of exact diagonalization while mitigating the fermionic sign or phase problem. Combined with the systematically improvable constrained-path~\cite{zhang1997constrained} or phaseless approximation~\cite{zhang2003quantum}, AFQMC has been successfully applied to a wide range of physical systems, including lattice models in condensed matter physics~\cite{xu2024coexistence}, \textit{ab initio} quantum chemistry~\cite{lee2020performance}, real materials~\cite{shi2021some}, and nuclear shell models~\cite{PhysRevLett.111.012502,bonnard2016constrained}. In many applications, it provides an attractive balance between accuracy and computational scalability.

For systems with strong pairing correlations, however, such as heavy open-shell nuclei near the neutron drip line~\cite{bohr1958possible}, the Slater determinant representation becomes inefficient. In contrast, a Hartree-Fock-Bogoliubov (HFB) wave function explicitly incorporates anomalous pairing fields, whereas representing the same paired state within the Slater determinant manifold generally requires an impractically large linear combination of determinants~\cite{TB:ring2004nuclear}. This observation has motivated substantial efforts to extend AFQMC from the Slater determinant manifold to the larger space of HFB wave functions~\cite{juillet2017phaseless,vitali2024monte,shi2017many}. Such an extension combines the ability of HFB theory to describe pairing correlations with the many-body accuracy of AFQMC. Nevertheless, developing an efficient, flexible, and numerically robust AFQMC framework for HFB wave functions remains an important challenge.

In our previous work, we introduced a stochastic gauge formalism for AFQMC in the Slater determinant manifold~\cite{mywork}. Originally inspired by Gaussian phase-space quantum Monte Carlo~\cite{corney2006gaussian,corney2005gaussian}, this formalism exploits stochastic gauge degrees of freedom to provide additional flexibility in controlling the stochastic evolution of walkers, thereby offering the potential to improve both sampling efficiency and accuracy. In the present work, we extend this formalism to the HFB manifold. In addition, we introduce a more flexible representation of HFB walkers by deriving an explicit analytical expression for the normalization factor associated with the representation matrix relative to the reference vacuum. Together, these developments establish a more flexible and robust AFQMC framework for HFB wave functions.

This paper is organized as follows. Section~\ref{Sec: AFQMC} briefly reviews the standard AFQMC formalism. Section~III presents its extension to the HFB manifold, including efficient propagation schemes for HFB walkers in both the product-state and Thouless-state representations, as well as the formulation of the stochastic generalized one-body propagator with stochastic gauge. Section~\ref{Sec: results} presents numerical benchmarks using the Richardson pairing Hamiltonian. Finally, conclusions and future perspectives are given in Section~\ref{Sec: con}.

\section{Review of AFQMC}\label{Sec: AFQMC}

In this section, we briefly review the AFQMC formalism on the Slater determinant manifold from a numerical perspective. More comprehensive reviews can be found in Refs.~\cite{lee2022twenty,shi2021some,motta2018ab}. 

As in other projection QMC methods, AFQMC projects an initial state onto the ground state through imaginary-time evolution,
\begin{align}
    \ket{\Psi_{\mathrm{gs}}} \propto \lim_{\tau \rightarrow \infty}e^{-\tau (\an H - E_0)} \ket{\Psi_0},
\end{align}
where $E_0$ is an estimate of the ground-state energy introduced to stabilize the numerical propagation and $\ket{\Psi_0}$ is an arbitrary initial state with nonzero overlap with the exact ground state and is typically determined from Hartree-Fock or density functional theory~\cite{lee2020utilizing,zhang2003quantum}. 

In a path-integral formulation, the projection is carried out through a sequence of short time steps. The small-time propagator is factorized using Trotter-Suzuki decomposition~\cite{trotter1959product} into one-body and two-body components, with the latter decoupled via a Hubbard-Stratonovich (HS) transformation~\cite{hubbard1959calculation}. For instance, assuming that the interaction term is of quadratic form $\an V = -\frac{1}{2}\lambda \an O^2$, we have
\begin{align}
    e^{-\tau \an V} = \int d\sigma p(\sigma)\exp(\sigma\sqrt{\tau}\sqrt{\lambda} \an O),
\end{align}
where $\sigma$ is the auxiliary field and $p(\sigma)$ is a Gaussian distribution. The many-body propagator is thereby reformulated as a high-dimensional path integral over one-body propagators parameterized by the auxiliary fields. The one-body propagator maps a Slater determinant onto another Slater determinant via Thouless's theorem~\cite{thouless1960stability, thouless1961vibrational}. The resulting integral is evaluated stochastically using Monte Carlo sampling, thereby mapping the exact dynamics onto a stochastic process over the Slater determinant manifold. 

To reduce statistical fluctuations, AFQMC employs an importance-sampling transformation based on a trial wave function $\ket{\Psi_T}$~\cite{zhang2003quantum}. This transformation substantially improves the efficiency and stability of the stochastic propagation. In this framework, the exact wave function can be reconstructed by
\begin{align}
    \ket{\Phi(\tau)} = \mathbb{E}\left[ \Omega(\tau)\frac{\ket{\Psi(\tau)}}{\iner{\Psi_T}{\Psi(\tau)}}\right],
\end{align}
where $\mathbb{E}[\cdot]$ is the ensemble average over Slater determinant walkers $\ket{\Psi}$, and $\Omega$ is a modified complex weight proportional to the overlap $\iner{\Psi_T}{\Psi}$. The resulting stochastic evolution of walkers and weights is given by
\begin{align}
    \ket{\Psi_{i+1}} = \an B(\sigma-\bar \sigma) \ket{\Psi_i},\quad
    \Omega_{i+1} = \Omega_ie^{-\tau (E_H - E_0)},
\end{align}
where $\bar \sigma$ is the force bias introduced to reduce variance~\cite{zhang2003quantum}
\begin{align}
    \bar \sigma = -\sqrt{\tau}\sqrt{\lambda} \frac{
    \mat{\Psi_T}{\an O}{ \Psi_i}
    }{
    \iner{\Psi_T}{\Psi_i}
    },
\end{align}
$\an  B(\sigma,\bar \sigma)$ is the stochastic one-body propagator obtained from a second-order Trotter decomposition,
\begin{align}
    \an B(\sigma - \bar \sigma) = e^{-\tau \an T/2}\exp\left( (\sigma - \bar \sigma)\sqrt{\tau}\sqrt{\lambda}\an O\right)e^{-\tau \an T/2},
\end{align}
and $E_H$ is the hybrid energy
\begin{align}
    E_H = -\frac{1}{\tau}{\rm log}\left(e^{-\frac{\bar \sigma^2}{2} + \sigma \bar \sigma}\frac{\iner{\Psi_T}{\Psi_{i+1}}}{\iner{\Psi_T}{\Psi_i}}\right).
\end{align}

In general, both the weights and orbitals are complex quantities. In the absence of constraints, the weights undergo an unconstrained random walk in the complex plane during the stochastic evolution, leading to a severe phase problem. To mitigate this issue, the phaseless approximation is proposed with an approximate trial wave function~\cite{zhang2003quantum}. This involves a positive-weight approximation and a cosine projection of the rotation angle
\begin{align}
    \Omega_{i+1}' = |\Omega_i| e^{-\tau ({\rm Re}(E_H) - E_0)}\times {\rm max}\left( 0, \cos(\Delta \theta)\right) 
\end{align}
with
\begin{align}
    \Delta \theta = \text{Im log}  \left(\frac{\iner{\Psi_T}{\Psi_{i+1}}}{\iner{\Psi_T}{\Psi_i}}\right).
\end{align}
The phaseless approximation maintains numerical stability while introducing a systematic bias, which is sensitive to the quality of trial wave function~\cite{shi2021some}.

Within the phaseless approximation, the ground-state energy is evaluated using the mixed estimator
\begin{align}
    E(\tau) = \frac{
    \mathbb{E}[\Omega(\tau)' {\rm Re}(E_L(\tau))]
    }{
    \mathbb{E}[\Omega(\tau)']
    },
\end{align}
where $E_L$ is the local energy 
\begin{align}
    E_L(\tau) = \frac{
    \mat{\Psi_T}{\an H}{\Psi(\tau)}
    }{
    \iner{\Psi_T}{\Psi(\tau)}
    },
\end{align}
which requires the calculation of overlaps and Green's functions between Slater determinants. The extension of this formalism to the HFB manifold therefore requires corresponding generalizations of the overlap, Green's function, and walker propagation while preserving the efficiency of the stochastic sampling.

\section{AFQMC in the HFB manifold}
As discussed in the preceding section, standard AFQMC relies on the stochastic evolution of walkers under one-body propagators that preserve the Slater determinant structure, thereby restricting the random walk to the Slater determinant manifold. Extensions of AFQMC to the more general Hartree-Fock-Bogoliubov (HFB) manifold have previously been developed~\cite{juillet2017phaseless,vitali2024monte,shi2017many}. In this section, we focus on the key ingredients of this formulation related to the propagation of HFB walkers while introducing two methodological improvements: a more flexible representation of HFB walkers and a reformulation of the stochastic propagator that incorporates stochastic gauge degrees of freedom. The corresponding expressions for overlaps and Green's functions in the HFB manifold are presented in the Appendix. With these ingredients in place, the AFQMC framework described in Sec.~II can be applied directly to the HFB manifold, with the walker $\ket{\Psi}$ replaced by an HFB wave function.

\subsection{Bogoliubov algebra}
\newcommand{\db}[1]{\dot{\bar{#1}}}
\newcommand{\ab}[1]{\an{\bar{#1}}}
We begin by reviewing the algebra of Bogoliubov quasiparticle operators and introducing the notation used throughout the remainder of this work.

Similar to the particle creation operators, which form a vector representation of one-body operators, Bogoliubov quasiparticle operators likewise form a vector representation of generalized one-body operators in the extended one-body space. For quasiparticle operators $\cra \beta_i$ and $\an{\bar \beta}_i$, defined via a (not necessarily unitary) Bogoliubov or canonical transformation $\mathcal{W}$~\cite{balian1969nonunitary}, we introduce the corresponding extended vectors,
\begin{align}
\ab B = 
    \begin{bmatrix}
        \an{\bar \beta} \\
         (\cra \beta)^T
    \end{bmatrix}
    ,\quad
    \cra{B}=
    \begin{bmatrix}
        \cra{\beta} & \an{\bar \beta}^T
    \end{bmatrix},
\end{align}
where $\cra \beta$ denotes the row vector of quasiparticle creation operators and $\an{\bar \beta}$ the column vector of quasiparticle annihilation operators. These vectors are related to the corresponding particle operators through the canonical transformation $\m W$,
\begin{align}
    \cra B = \cra C \mathcal{W}, \quad \an{\bar B} = \mathcal{W}^{-1} \an C,
\end{align}
where $\cra C$ and $\an C$ denote the extended vectors constructed from the particle operators $\cra c$ and $\an c$ associated with an orthonormal single-particle basis of dimension $M$. The quasiparticle operators satisfy the canonical anticommutation relations
\begin{align}\label{quasi-vector anti}
    \{\an{\bar B}, \cra B\} = \mathcal{I}, \quad \{\an{\bar B}, \an{\bar B}^T\} = \{(\cra{ B})^T, \cra B\} =\sigma,
\end{align}
where $\mathcal{I}$ is the identity matrix and $\sigma$ is the permutation matrix in the extended $2M$-dimension space, satisfying $\sigma^T = \sigma$ and $\sigma^2 = \mathcal{I}$. These relations are preserved by the transformation $\m W$ that satisfies $\sigma \m W^T\sigma = \m W^{-1}$ and belongs to the canonical transformation group. Note that $\mathcal{W}$ is not necessarily unitary; consequently, $\cra \beta$ and $\an{\bar \beta}$ are not Hermitian conjugates.

We define a generalized one-body operator as $\an K = \frac{1}{2}\cra B K \an{\bar B}$ where $K$ is an antisymmetric matrix ($\sigma K^T \sigma = -K$) in the extended one-body space. Using the anticommutation relations in Eq.~(\ref{quasi-vector anti}), one readily obtains
\begin{align}
    [
        \an K , \cra B 
    ]= \cra B K.
\end{align}
It follows that the generalized one-body propagator $\an F = \exp(\an K)$ induces a canonical transformation $F=\exp(K)$, under which the quasiparticle operators transform according to
\begin{align}\label{tran}
    \an F \cra B \an F ^{-1} = \cra B F,  \quad
    \an F \an{\bar B} \an F ^{-1} = F^{-1} \an{\bar B}.
\end{align}
Thus, $\cra B$ and $\an{\bar B}$ transform as representations of the generalized one-body operator in the extended one-body space. The canonical transformation $F$ takes the block form shown below,
\begin{align}
    F = \begin{bmatrix}
        \overline{U}&V^* \\
        \overline{V} &U^*
    \end{bmatrix},
\end{align}
where $U, V$ are $M\times M$ matrix satisfying, according to Eq.~(\ref{quasi-vector anti}), the canonical orthogonality conditions
\begin{align}\label{orth.}
    \overline{V}^TV^* + \overline{U}^TU^* =&  I,\nonumber\\
    U^\dagger V^* + V^\dagger U^* =& 0.
\end{align}
In the special case where $F$ is unitary, we have $\overline{U} = U$ and $\overline{V} = V$. 

\subsection{Propagation of HFB wave function}
Having established the underlying algebra and notation, we now turn to the numerical propagation of walkers in the Hartree-Fock-Bogoliubov (HFB) manifold. The propagation $e^{\an S}$ is generated by the generalized one-body operator $\an S = \frac{1}{2} \cra C S \an C$ where $S$ is a complex $2M \times 2M$ matrix satisfying the antisymmetry condition $\sigma S^T \sigma = -S$. As discussed in the previous subsection, the corresponding matrix representation $e^S$ is a generally non-unitary Bogoliubov transformation, which can be written in block form, 
\begin{align}
    e^{S} = 
    \begin{bmatrix}
    \overline{U}_s & V_s^* \\
    \overline{V}_s & U_s^*
    \end{bmatrix},
\end{align}
where the $U,V$ matrix satisfy the canonical orthogonality relations in Eq.~(\ref{orth.}). Our objective is to develop a numerically efficient implementation of the HFB walker propagation,
\begin{align}
    \ket{\Psi_{i+1}} = e^{\an S} \ket{\Psi_i},
\end{align}
where, according to Thouless' theorem, $\ket{\Psi_{i+1}}$ remains an HFB quasiparticle vacuum. As in previous HFB-AFQMC formulations~\cite{shi2017many}, the HFB walker may be represented either as a product state or in Thouless form. In this work, we develop a more flexible and robust formulation of both representations by allowing an arbitrary choice of reference vacuum and deriving the corresponding normalization factor directly from the representation matrix.

\subsubsection{Product state representation}
Similar to a Slater determinant, an HFB walker may be represented in an unnormalized product-state form as
\begin{align}
    \ket{\Psi_i^P} = \prod_{k = 1}^M \an \beta^{(i)}_k \ket{\Psi_0},
\end{align}
where $\ket{\Psi_0}$ is a reference quasiparticle vacuum associated with quasiparticle operators $\cra B_{0} = [\cra \beta{}^{(0)}, (\an \beta^{(0)})^T]$, which are defined through an unitary Bogoliubov transformation $\mathcal{W}_0$,
\begin{align}
    \cra B_0
     =
     \cra C
    \mathcal{W}_0 =
    \begin{bmatrix}
        \cra c & \an c^T
    \end{bmatrix}
    \begin{bmatrix}
        U_0 & V^*_0 \\
        V_0 & U^*_0
    \end{bmatrix}.
\end{align}
A common choice for the reference state is the particle vacuum annihilated by the operators $\an c$, corresponding to $\an \beta^{(0)} = \an c$. Unlike previous formulations, however, the present representation allows an arbitrary choice of reference vacuum. This flexibility effectively shifts the origin of the $U$- and $V$-matrix evolution and can improve numerical efficiency with a suitable vacuum such as the self-consistent HFB solution. In this representation, the quasiparticle operators are expressed as
\begin{align}
    \cra B_i = \begin{bmatrix}
        \cra \beta {}^{(i)} & (\an \beta^{(i)})^T
    \end{bmatrix}
     =
     \cra B_0
    \begin{bmatrix}
        U_i & V^*_i \\
        V_i & U^*_i
    \end{bmatrix}.
\end{align}
Using the transformation law in Eq.~(\ref{tran}), the propagated quasiparticle operators are obtained as
\begin{align}
    (\an \beta^{(i+1)})^T = e^{\an S} (\an \beta^{(i)})^T e^{ - \an S} = \cra B_0 \begin{bmatrix}
        V^*_{i+1} \\ U^*_{i+1}
    \end{bmatrix}
\end{align}
with
\begin{align}\label{24}
    \begin{bmatrix}
        V^*_{i+1} \\ U^*_{i+1}
    \end{bmatrix}
    =\mathcal{W}_0^\dagger e^S \mathcal{W}_0 
    \begin{bmatrix}
        V^*_{i} \\ U^*_{i}
    \end{bmatrix}
    =
    \begin{bmatrix}
        \bar{\mathcal{U}}_s & \mathcal{V}_s^* \\
        \bar{\mathcal{V}}_s & \mathcal{U}_s^*
    \end{bmatrix}
    \begin{bmatrix}
        V^*_{i} \\ U^*_{i}
    \end{bmatrix}.
\end{align}
In principle, from the group structure of canonical transformation, the transformed operators satisfy the anticommuting relation $\{\an \beta^{(i+1)}_k, \an \beta^{(i+1)}_j\} =0$, implying that the skew symmetry of $U_{i+1}^T V_{i+1}$ is maintained throughout the propagation. Finite-precision arithmetic, however, introduces round-off errors that may violate this skew-symmetry condition. In addition, since the transformation is in general not unitary, the orthogonal relation $V^T_{i+1} V^*_{i+1} + U^T_{i+1} U_{i+1}^*=I$ may be broken. To restore the canonical structure and improve numerical stability, we should first enforce the skew-symmetry condition and subsequently reorthogonalize the walker using, for instance, a Gram–Schmidt procedure.

Finally, the transformed walker in a product form is written as
\begin{align}
    \ket{\Psi_{i+1}} = e^{\an S} \ket{\Psi_i^P} = \mathcal{N} \prod_k \an \beta_k^{(i+1)} \ket{\Psi_0}=\mathcal{N} \ket{\Psi_{i+1}^P},
\end{align}
where $\mathcal{N}$ is a normalization factor required for evaluating the hybrid energy and imposing the phaseless constraint. The normalization factor is defined by
\begin{align}
    \mathcal{N} = \frac{
    \mat{\Phi}{e^{\an S}}{  \Psi_i^P}
    }{
    \iner{\Phi}{\Psi_{i+1}^P}
    },
\end{align}
where $\Phi$ is an arbitrary wave function not orthogonal to $\ket{\Psi_{i+1}^P}$. Choosing $\ket{\Phi} = \ket{\Psi_0}$, the normalization factor can be expressed analytically in terms of the $U,V$ matrix, as derived in the Appendix,
\begin{align}
    \mathcal{N}= \sqrt{\det{(\mathcal{U}_s^*)}} \frac{
    {\rm pf}( U_{i+1}^\dagger (\mathcal{U}_s^\dagger)^{-1} V_i^* )
    }{
    {\rm pf}(U_{i+1}^\dagger  V_{i+1}^*)
    }.
\end{align}
When the walker becomes nearly orthogonal to the reference vacuum, the above expression may become numerically unstable. In this case, it is advantageous to choose $\ket{\Phi} = \ket{\Psi_i^P}$, from which, together with Eqs.~(\ref{olp_prod_1}) and (\ref{c2}), one obtains
\begin{align}
    \mathcal{N}=\sqrt{\det(U^*)} \frac{{\rm pf}(S^{(i,i)})}{{\rm pf}(S^{(i,i+1)})},
\end{align}
where
\begin{align}
    U^* 
    = V_i^T  V_{i+1}^* + U_i^T  U_{i+1}^* 
\end{align}
and
\begin{align}
    S^{(i,j)} =& \begin{bmatrix}
        V_i^TU_i & V_i^T  V_{j}^* \\
        - V_{j}^*V_i &  U_{j}^\dagger  V_{j}^*
    \end{bmatrix}.
\end{align}
The overlap and generalized density matrices required for evaluating observables through Wick's theorem can likewise be expressed entirely in terms of the $U$- and $V$-matrices, as shown in the Appendix. Consequently, the complete AFQMC simulation in the product-state representation is fully parametrized by the evolution of the $U$- and $V$-matrices. Since the reference vacuum defines the origin of this parametrization, it may be chosen freely to suit the physical system under consideration.

\subsubsection{Thouless representation}
An HFB wave function that is not orthogonal to the reference vacuum can alternatively be represented in the unnormalized Thouless form,
\begin{align}
    \ket{\Psi_i^T} = \exp\left(\frac{1}{2}\cra \beta {}^{(0)} Z_i \an \beta^{(0)}\right)\ket{\Psi_0}, 
\end{align}
where $Z_i=( V_iU_i^{-1})^*$ is a $M\times M$ skew-symmetric matrix. When the reference state $\Psi_0$ is chosen as the particle vacuum, the Thouless representation describes fully paired states with nonzero overlap with the vacuum. Allowing an arbitrary reference vacuum significantly extends the applicability of the Thouless representation by enabling the description of states containing unpaired fermions, which are essential for systems with odd particle number or spin polarization.

The propagated HFB walker can be written in the same Thouless form,
\begin{align}
    \ket{\Psi_{i+1}} = \mathcal{N} \exp\left(\frac{1}{2}\cra \beta {}^{(0)} Z_{i+1} \an \beta^{(0)}\right)\ket{\Psi_0} = \mathcal{N} \ket{\Psi_{i+1}^T},
\end{align}
where the propagation is completely characterized by the updated Thouless matrix $Z_{i+1}$ and the normalization factor $\m N$. The latter is required for evaluating the hybrid energy and imposing the phaseless constraint. While the Thouless propagation itself is well established, an explicit analytical expression for the normalization factor in terms of the Thouless matrix is essential for AFQMC simulations. By introducing a pair of biorthogonal Fock states, we derive in the Appendix the compact expression
\begin{align}
    \mathcal{N} = \mat{\Psi_0}{e^{\an S}}{ \Psi_i} = \sqrt{\det{(\overline{\m V}_s Z_i + \m U_s^*)}}.
\end{align}
The Thouless matrix evolves according to
\begin{align}
    Z_{i+1}
    = V_{i+1}^* (U_{i+1}^*)^{-1}
    =\left(\overline{\m U}_sZ_i + V_s^*\right) \left(\overline{\m V}_s Z_i + \m U_s^*\right)^{-1}.
\end{align}
As shown in the Appendix, the overlap, generalized density matrices, and the normalization factor can all be expressed analytically in terms of the Thouless matrix $Z$. Consequently, the AFQMC propagation can be carried out solely in the space of $Z$, whose dimension is approximately half that of the product-state representation based on the $U,V$ matrix. The Thouless representation is therefore expected to offer improved computational efficiency. In addition, the $Z$ matrix is invariant under the reorthogonalization of the $U,V$ matrix. As a result, the propagation is unaffected by the orthogonalization procedure, and numerical stabilization only requires enforcing the skew-symmetry of $Z$. This leads to a simpler and potentially more stable propagation scheme. The Thouless representation, however, is restricted to the subset of HFB states that are nonorthogonal to the chosen reference vacuum. Consequently, the choice of reference vacuum plays a crucial role in both the efficiency and the robustness of the numerical simulation. When the overlap with the reference vacuum becomes vanishingly small and the Thouless representation is no longer well defined, the product-state representation should be employed instead.

\subsection{Imaginary-time propagator with stochastic gauge}
In the previous subsection, we established a numerically robust and flexible framework for propagating HFB walkers under a given generalized one-body propagator. We now turn to the construction of this propagator by extending our previously developed stochastic gauge formalism to the HFB manifold.

In conventional AFQMC, the imaginary-time propagator is factorized using the Hubbard-Stratonovich (HS) transformation, which decouples the two-body interaction into stochastic generalized one-body propagators. This formulation is typically combined with a force bias to reduce both statistical fluctuations and the bias introduced by the phaseless approximation. In our previous work~\cite{mywork}, we showed that the stochastic decoupling possesses additional stochastic gauge degrees of freedom, providing greater flexibility in the construction of the stochastic propagator on the Slater determinant manifold. Although different gauge choices yield equivalent exact imaginary-time evolution, they can exhibit different numerical efficiencies and, in the presence of the phaseless approximation, different systematic biases. In this section, we show that the stochastic gauge formalism can be naturally extended to the HFB manifold.

Without loss of generality, we write the Hamiltonian as 
\begin{align}\label{ham_qmc}
    \an H = \an{\mathcal{T}} - \frac{1}{2}  \lambda_i \an{\mathcal{O}}_i^2 +c,
\end{align}
where $\an{\m T} = \cra C \m T \an C /2$ and $\an{\m O} = \cra C \m O \an C /2$ are generalized one-body operators, $c$ is a constant, and repeated indices imply summation unless otherwise specified.

\subsubsection{Drift gauge}
The interaction component of the short-time imaginary-time propagator can be reformulated by introducing a local background field $g_i$, referred to as the drift gauge in our previous work~\cite{mywork}. Applying the Hubbard-Stratonovich transformation to the shifted interaction yields
\begin{align}\label{39}
    e^{\frac{\tau}{2}\lambda_i \an{\m O}^2_i} = \int d\sigma p(\sigma) e^{-\frac{\tau}{2}\lambda_i g_i^2+ \sigma_i g_i\sqrt{\tau} \sqrt{\lambda_i}}\an B_V(\sigma_i,g_i) + O(\tau^2),
\end{align}
where $\sigma$ is a vector of auxiliary field, $p(\sigma)$ is the standard Gaussian measure, and $\an B_V(\sigma_i, g_i)$ is the stochastic propagator associated with the chosen drift gauge,
\begin{align}
    \an B_V(\sigma_i, g_i) = \exp\left( - \tau
\lambda_ig_i \an{\mathcal{O}}_i+ \sigma_i  \sqrt{\tau}\sqrt{\lambda_i} \an{\mathcal{O}}_i
\right).
\end{align}
The drift gauge introduces additional freedom in the stochastic decomposition while leaving the exact imaginary-time propagator unchanged. As discussed in Ref.~\cite{mywork}, different choices of $g_i$  redistribute stochastic fluctuations between the walker evolution and the walker weights, thereby mediating a trade-off between orbital drift and weight diffusion. In particular, choosing $g_i = -\av{\an{\m O}_i}_{T,s}$, where $\av{\cdot}_{T,s}$ denotes the mixed estimator between the trial wave function and the walker, minimizes the weight diffusion and recovers the conventional force-bias formulation of AFQMC.

\subsubsection{Fermi gauge}
In addition, we can also restrict the HS transformation to the residual interaction. In our previous work, this construction was introduced as the Fermi gauge~\cite{mywork}. It is achieved by first expressing the interaction in normal-ordered form with contractions defined between the correlated trial wave function and the walker. The short-time propagator can then be approximated as
\begin{align}
e^{
\frac{1}{2}\tau \lambda_i \an{\mathcal{O}}_i^2 
}
=
e^{K^{(0)}} :\exp\left(\an K^{(1)} + \frac{1}{2}\tau \lambda_i \an{\mathcal{O}}_i^2
\right): + O(\tau^2),
\end{align}
where $K^{(0)}$ is a scalar term and $\an K^{(1)}$ is a generalized one-body operator. Both quantities are obtained by matching the Taylor expansions of the two sides using Wick's theorem. In principle, the normal-ordered approximation can be systematically improved by retaining higher-order terms in the expansion, although this leads to a more complicated formulation. At first order, $K^{(0)}$ and $\an K^{(1)}$ arise solely from the interaction term $\an{\mathcal{O}}_i^2 $ and are given by
\begin{align}\label{40}
    K^{(0)} =& \tau \frac{\lambda_i}{8}{\rm Tr}(\mathcal{O}_iR)^2+ \tau \frac{\lambda_i}{4}{\rm Tr}(\mathcal{O}_iR\mathcal{O}_iG) + O(\tau^2),\nonumber \\
    K^{(1)} =& \frac{1}{2}\tau \lambda_i{\rm Tr}(\mathcal{O}_iR) \mathcal{O} - \frac{1}{2}\tau \lambda_i\mathcal{O}_i(R - G)\mathcal{O}_i+ O(\tau^2),
\end{align}
where ${\rm Tr}(\cdot)$ denotes the trace operation in the extended one-body space, $R$ is the generalized one-body density matrix between the trial wave function and walker, and $G = \sigma R^T \sigma = I - R $ is the `abnormal' density matrix. They satisfy $GR = RG = 0$ and $R^2 = R$, irrespective of whether the trial state is a single quasiparticle vacuum or a superposition of quasiparticle vacua, as shown in the Appendix. One readily verifies that the one-body operator $\an K^{(1)}$ in Eq.~(\ref{40}) has the form of the conventional HFB mean-field Hamiltonian~\cite{TB:ring2004nuclear}. For this reason, we also refer to the Fermi gauge as the mean-field gauge in the present work.

Applying the HS transformation to the normal-ordered propagator yields
\begin{align}
    e^{
\frac{1}{2}\tau \lambda_i \an{\mathcal{O}}_i^2 
} =& \int d\sigma  p(\sigma) e^{K^{(0)}} \times \nonumber \\&:\exp\left(\an K^{(1)} 
+ \sigma_i\sqrt{ \tau} \sqrt{\lambda_i}\an{\mathcal{O}}_i
\right): + O(\tau^2).
\end{align}
Since the walker is the right vacuum of the normal-ordered quasiparticle operators, the stochastic propagation is restricted to the residual particle-hole excitation channels. Compared with the drift-gauge formulation in Eq.~(\ref{39}), the HS transformation is now performed exactly on the normal-ordered interaction. Consequently, the leading time-discretization error originates from the normal-ordering approximation rather than from the HS transformation itself~\cite{sukurma2024toward}. In addition, since the HFB walker explicitly breaks the abelian $U(1)$-symmetry, the symmetry gauge introduced in the previous work~\cite{mywork} can also be incorporated to improve the sampling efficiency for $U(1)$-conserving Hamiltonian.

\section{Illustration} \label{Sec: results}
\begin{figure}[t]
\centering
\hspace*{-1.0cm}
\includegraphics[width=0.45\textwidth]{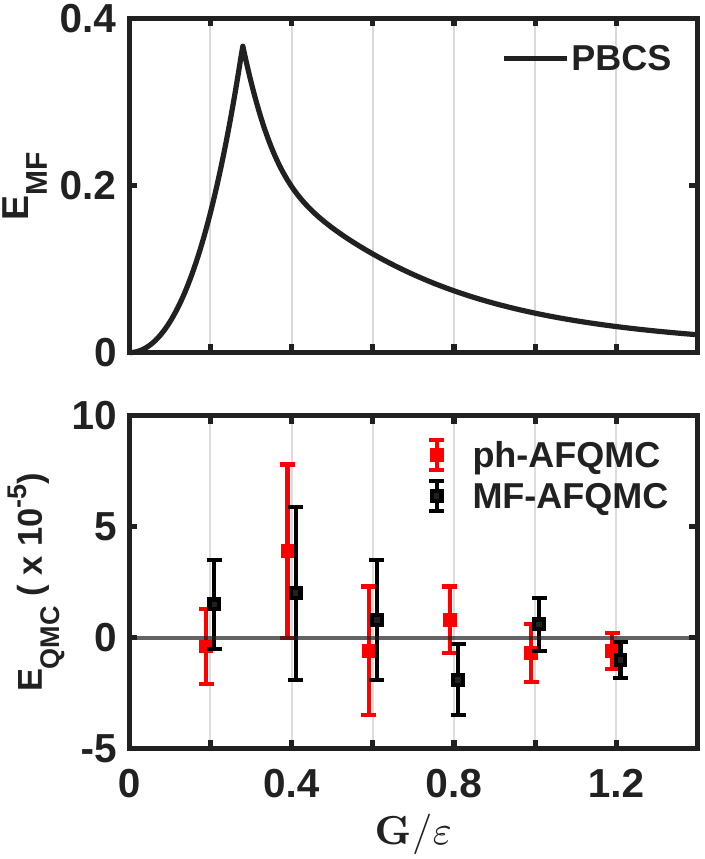}
    \caption{Deviation of the calculated ground-state energy from the exact solution for the Richardson model with $N=10$ equally spaced doubly degenerate single-particle levels at half filling. Upper panel: Particle-number-projected BCS (PBCS) results. Lower panel: Conventional phaseless AFQMC (ph-AFQMC) and the mean-field-gauge AFQMC (MF-AFQMC), both employing the PBCS wave function as the trial state and $\mu = -G/2$ as the chemical potential. The energy is expressed in units of the level spacing $\varepsilon$.}
    \label{fig:2}
\end{figure}

As an illustrative example, we consider the Richardson model, which describes a pure pairing Hamiltonian with contact interactions between fermions occupying $N$ doubly degenerate equally spaced single-particle levels,
\begin{align}
    \an H = \sum_{i= 1}^N \varepsilon_i\sum_{\sigma = \uparrow \downarrow}\cra c_{i \sigma} \an c_{i \sigma}  - G \sum_{ij} \cra c_{i \uparrow} \cra c_{i \downarrow} \an c_{j \downarrow} \an c_{j \uparrow},
\end{align}
where the single-particle energies are given by $\varepsilon_i = (i-\frac{N+1}{2})\varepsilon$, with $\varepsilon$ denoting the level spacing. Originally introduced as a simplified model of nuclear superfluidity, the Richardson model provides an accurate description of pairing correlations in nuclei such as the Sn and Pb isotopic chains~\cite{de2014probing,richardson1963application}. Owing to its exact analytical solution\cite{richardson1964exact,richardson1965number}, it has also become a standard benchmark for assessing the accuracy of approximate many-body methods~\cite{companys2024eigenvector,rigo2023solving,brolli2025diagrammatic,hu2025ab,hu2026stochastic}. Moreover, the dominant pairing correlations are well described at the mean-field level by BCS theory, which predicts a transition from the normal to the superconducting phase at a critical pairing strength, as illustrated in Figure~\ref{fig:2}. These features make the Richardson model an ideal testing ground for the present HFB-AFQMC formulation.

\begin{figure}[t]
\centering
\hspace*{-1.0cm}
\includegraphics[width=0.45\textwidth]{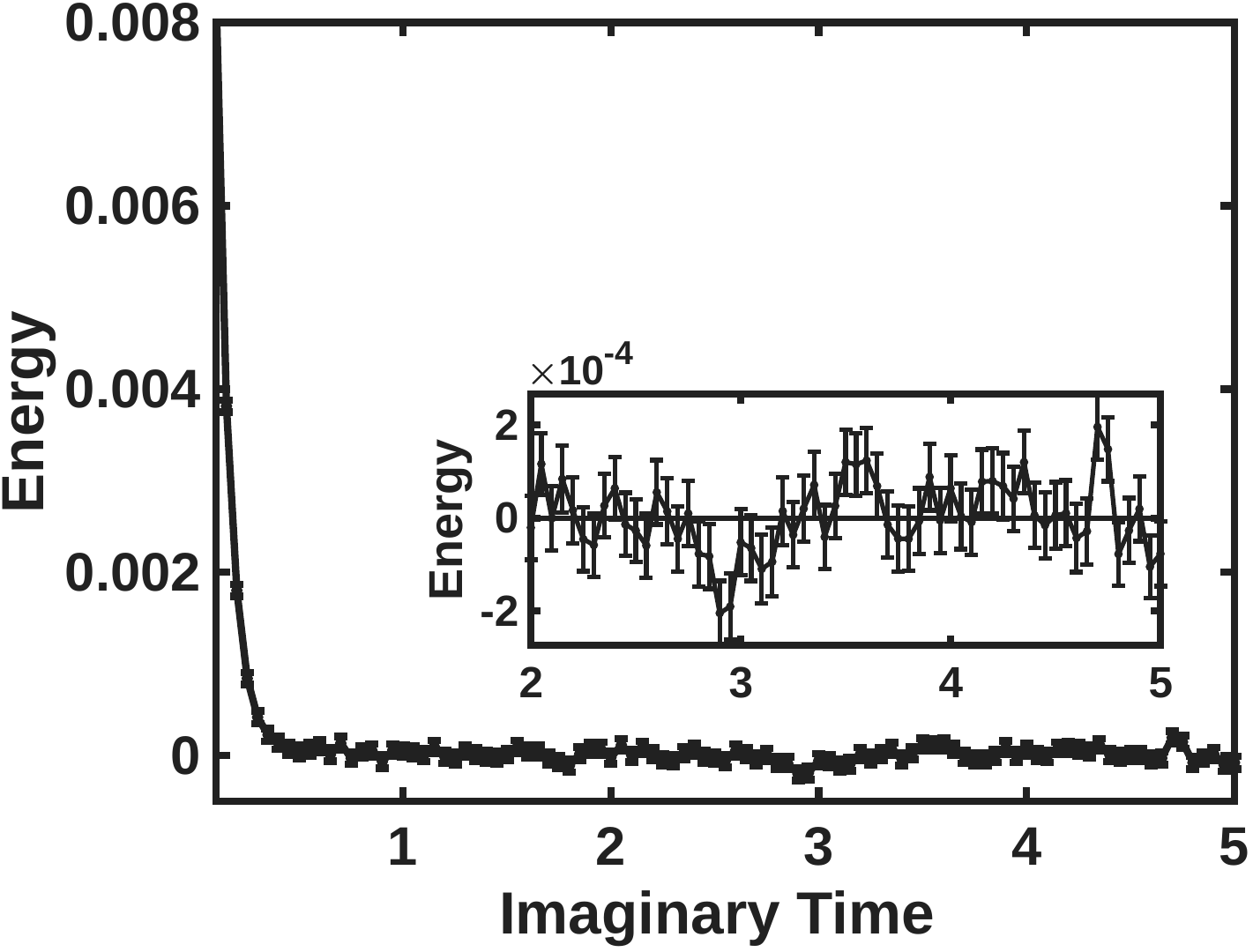}
    \caption{Imaginary-time evolution of the projected energy for the Richardson model with $N=10$ equally spaced doubly degenerate single-particle levels at half filling and attractive pairing strength $G/\varepsilon=1$. The projected energy is evaluated in MF-AFQMC using the mixed estimator with a particle-number-projected BCS trial wave function and a chemical potential $\mu=-G/2$. The energy is shown relative to the exact ground-state energy obtained by exact diagonalization. Energies are expressed in units of the level spacing $\varepsilon$, and imaginary time is expressed in units of $\hbar/\varepsilon$. The inset enlarges the converged region to show the statistical uncertainties.}
    \label{fig:1}
\end{figure}
In the present calculations, the pairing interaction is decoupled using a continuous Hubbard-Stratonovich transformation after rewriting it in the quadratic form $-G (\an S_x^2 + \an S_y^2 )$, where
\begin{align}
    \an S_x =\frac{1}{2} \sum_p \cra c_{p\uparrow} \cra c_{p\downarrow} + \an c_{p\downarrow} \an c_{p \uparrow},\nonumber \\
    \an S_y = \frac{1}{2i}\sum_p \cra c_{p\uparrow} \cra c_{p\downarrow} - \an c_{p\downarrow} \an c_{p \uparrow}. 
\end{align}
The resulting stochastic one-body propagator breaks the $U(1)$ symmetry and corresponds to a special Bogoliubov transformation that preserves the pairing structure. Consequently, the general HFB formalism developed in the preceding sections reduces naturally to a more simplified BCS manifold, where the BCS walkers are conveniently parametrized by the Bogoliubov amplitudes $u, v$ in the product state representation. The corresponding auxiliary fields introduce purely imaginary stochastic fluctuations, making the phaseless approximation essential for maintaining numerical stability. For numerical simulation, unless otherwise stated, the self-consistent BCS solution is adopted both as the initial walker and as the reference vacuum for the walker representation. To restore the broken $U(1)$ symmetry and guide the stochastic evolution, we employ the particle-number-projected BCS (PBCS) state as the trial wave function. As illustrated in Figure~\ref{fig:2}, the PBCS state recovers most of the correlation energy, especially in the strong-pairing regime. The corresponding Ito-stochastic differential equations governing the quasiparticle evolution are also provided in the Appendix for completeness.

\begin{figure}[b]
\centering
\hspace*{-1.0cm}
\includegraphics[width=0.45\textwidth]{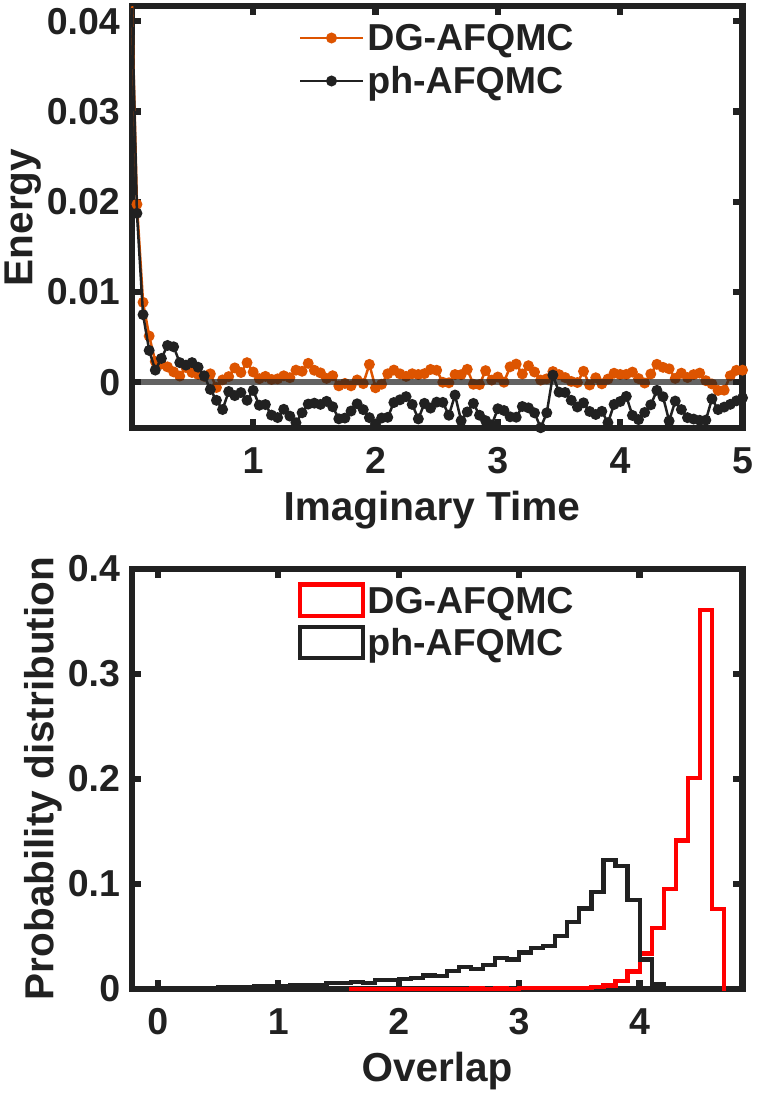}
    \caption{Upper panel: Imaginary-time evolution of the energy deviation from the exact ground-state energy for the Richardson model at $G/\varepsilon=1$ with the chemical potential $\mu = 0$. Results obtained with the conventional phaseless AFQMC (ph-AFQMC) and the drift-gauge AFQMC (DG-AFQMC) are compared. Lower panel: Probability distributions of the magnitude of the overlap between the trial wave function and the walkers. The drift gauge suppresses the heavy tail in the small-overlap region, leading to improved accuracy under the phaseless approximation.}
    \label{fig:3}
\end{figure}
Figure~\ref{fig:1} presents the imaginary-time evolution of the energy obtained with AFQMC formulated using the mean-field gauge introduced in Eqs.~(40) and (41), hereafter referred to as MF-AFQMC. The projected energy converges smoothly toward the exact ground-state value during the imaginary-time evolution, while the statistical uncertainties remain well controlled throughout the propagation, demonstrating the numerical stability of the proposed formulation. Figure~\ref{fig:2} further compares the ground-state energies obtained with MF-AFQMC and the conventional phaseless AFQMC employing the standard force bias over a range of pairing strengths. Both methods reproduce the exact ground-state energies within statistical uncertainties across the entire interaction range, demonstrating that the proposed formulation preserves the accuracy of the conventional approach.

As shown in the Appendix, the conventional phaseless AFQMC may be interpreted as a stochastic Brownian motion of quasiparticle states driven by the standard BCS mean-field Hamiltonian. In contrast, the mean-field gauge incorporates additional Hartree-Fock contributions into the deterministic drift while restricting the stochastic propagation to the residual interaction. Nevertheless, both formulations exhibit comparable accuracy for the Richardson model. Although not shown here, we also examined a more challenging scenario by employing a normal-phase trial wave function in the strong-pairing regime, where the phaseless approximation is expected to introduce a larger systematic bias. Even under these more demanding conditions, the two methods yield nearly identical results. These observations suggest that the Richardson model is not sufficiently sensitive to expose a clear advantage of the mean-field gauge.

To illustrate the effect of the drift gauge, we deliberately tune the chemical potential to a more challenging regime in which the overlap distribution develops a pronounced heavy tail toward the small-overlap region, as shown in Figure~\ref{fig:3}. In this regime, the drift term of the conventional AFQMC becomes increasingly singular as the overlap approaches zero, producing rare walker configurations with exceptionally large stochastic fluctuations. These events are subsequently suppressed by the cosine projection in the phaseless approximation to maintain numerical stability, at the expense of an increased systematic bias. To regularize the drift, we employ the drift gauge introduced in Eq.~(38) by choosing $g_i = -\av{\an{\m O}_i}_{s,s}$ where $\av{\cdot}_{s,s}$ denotes the expectation value evaluated with respect to the walker. As discussed in the Appendix, this choice minimizes the diffusion of the walker trajectories, although it generally increases the fluctuations of the walker weights. As shown in Figure~\ref{fig:3}, the drift gauge substantially suppresses the heavy tail of the overlap distribution, resulting in a more compact distribution and enabling the ground-state energy to be reproduced with high accuracy. 

Although the Richardson model provides a valuable benchmark for validating both the HFB formulation and the stochastic gauge extension, it remains a highly simplified pairing Hamiltonian. Consequently, neither the advantages of the mean-field gauge nor those of the Thouless representation become pronounced in the present calculations. For more realistic Hamiltonians, where the mean-field structure and residual interactions are considerably more intricate, incorporating additional mean-field contributions into the deterministic propagation may lead to a greater reduction of residual stochastic fluctuations and the associated phaseless bias. Likewise, the compact parametrization and improved numerical stability of the Thouless representation are expected to become more beneficial in large-scale simulations. The encouraging performance of the drift gauge in the challenging test case presented above further supports the potential of these developments for realistic nuclear many-body calculations.

\section{Conclusion}\label{Sec: con}
In this work, we have formulated and implemented the phaseless auxiliary-field quantum Monte Carlo method in the Hartree-Fock-Bogoliubov (HFB) manifold for systems with strong pairing correlations. The formulation extends the conventional Slater-determinant-based AFQMC framework to quasiparticle vacua while preserving the general structure of the phaseless algorithm. In particular, we introduced a flexible representation of HFB walkers in both the product-state and Thouless representation with an explicit normalization factor derived from the representation matrix relative to an arbitrary reference vacuum. Building upon our previous work on stochastic gauge freedom in the Slater determinant manifold, we further extended the stochastic gauge formalism to the HFB framework. The resulting generalized one-body propagator naturally incorporates both the drift gauge and the mean-field (Fermi) gauge, providing additional flexibility in constructing the stochastic evolution of HFB walkers. The formulation is sufficiently general to accommodate arbitrary Bogoliubov transformations and can be integrated directly into the standard phaseless AFQMC algorithm. As a proof of concept, we benchmarked the method using the Richardson pairing model. The calculations demonstrate that the HFB-AFQMC framework is numerically stable and accurately reproduces the exact ground-state energy over a broad range of pairing strengths. For this model, the proposed mean-field gauge yields accuracy comparable to that of the conventional force-bias formulation, even in regimes where the phaseless approximation introduces a relatively large systematic bias. In contrast, the drift gauge substantially improves the overlap distribution in a deliberately challenging test case, illustrating its potential for reducing the occurrence of unstable stochastic trajectories. Although the Richardson model provides a useful benchmark for validating the present formulation, it remains a highly simplified pairing Hamiltonian. The advantages of the mean-field gauge and the Thouless representation are therefore not expected to be fully manifested in such a model. More realistic nuclear Hamiltonians, where the mean-field structure and residual interactions are considerably richer, provide a more suitable setting for assessing their practical benefits. The present formulation establishes the theoretical and computational foundation for such applications. Future work will focus on realistic nuclear shell-model Hamiltonians, where the interplay between pairing, deformation, and symmetry restoration is expected to provide a more stringent test of the HFB-AFQMC framework and its associated stochastic gauge techniques.  

\section*{Acknowledgments}
The author thanks Haozhao Liang for his guidance. This research was supported in part by Forefront Physics and Mathematics Program to Drive Transformation (FoPM), a World-leading Innovative Graduate Study (WINGS) Program, the University of Tokyo.

\appendix
\section{Properties of generalized one-body density matrix}
In this appendix, we summarize several properties of the generalized one-body density matrix that are used throughout the main text. As discussed in Sec.~III A, Bogoliubov quasiparticle operators transform as vector representations of generalized one-body operators in the extended $2M$-dimensional one-body space. To facilitate the discussion, we introduce the vector states $\ket{b}$ and $\ket{\dot b}$ associate with quasiparticle operators $\cra b$ and $\an{b}$ that are assumed to be related by hermitian conjugation and also to satisfy the standard fermionic anticommutation relations. They form a closure relation in the extended one-body space
\begin{align}
    \m I = \sum_{i=1}^M\ket{b_i}\bra{b_i} + \ket{\dot b_i}\bra{\dot b_i}.
\end{align}
This vector representation provides a convenient and basis-independent formulation of the generalized one-body density matrix.

We first consider the generalized one-body density matrix defined between two nonorthogonal quasiparticle vacua,
\begin{align}
    R_{k, l} = \frac{\mat{0_a}{\cra{C}_{l} \an{C}_{k}}{0_b}}{\iner{0_a}{0_b}},\quad k,l=1,...,2M
\end{align}
where $\ket{0_{a,b}}$ are quasiparticle vacua associated with the quasiparticle operators $\cra A = [\cra a, \an a^T]$ and $\cra B = [\cra b, \an b^T]$. For simplicity, we assume that the quasiparticle operators are defined by unitary canonical transformations $\cra A = \cra C W_a$ and $\cra B = \cra C W_b$ with $W_a^\dagger = W_a^{-1}$ and $W_b^\dagger = W_b^{-1}$. Analogous to the transition density matrix between Slater determinants~\cite{PhysRev.101.1730}, the generalized density matrix admits the compact vector representation
\begin{align}
    R = \sum_{kl} \ket{\dot{ b}_k}f^{-1}_{kl}\bra{\dot a_l},
\end{align}
where $f_{ij} = \iner{\dot a_i}{\dot b_j}$ is an inner product in the extended one-body space. It immediately follows that $R$ is a projection operator satisfying $R^2 = R$, although it is generally non-Hermitian. Expressing the density matrix in the particle basis associated with the operators $\cra C$ yields
\begin{align}
    R=& \left[ 
    \begin{array}{cc}
        V_b^* \\
        U_b^*
    \end{array}
    \right](f)^{-1}\left[ 
    \begin{array}{cc}
        V_a^T &U_a^T
    \end{array}
    \right ] \nonumber \\
    =& \left[ 
    \begin{array}{cc}
        V_b^* \\
        U_b^*
    \end{array}
    \right](V_a^T V_b^* + U_a^T U_b^*)^{-1}\left[ 
    \begin{array}{cc}
        V_a^T &U_a^T
    \end{array}
    \right ] \nonumber \\
    =& \left[ 
    \begin{array}{cc}
        \rho &\kappa \\
        -\bar{\kappa}^* &1 - \rho^T
    \end{array}
    \right ],
\end{align}
from which the normal and anomalous density matrices are identified as
\begin{align}
    \rho_{ij} =& \frac{
    \mat{0_a}{\cra c_j \an c_i}{0_b}
    }{
    \iner{0_a}{0_b}
    } = (V_b^*(f)^{-1}V_a^T)_{ij},\nonumber \\
    \kappa_{ij} =& \frac{
    \mat{0_a}{\an c_j \an c_i}{0_b}
    }{
    \iner{0_a}{0_b}
    } = (V_b^*(f)^{-1}U_a^T)_{ij},\nonumber \\ 
    -\bar\kappa_{ij}^* =& \frac{
    \mat{0_a}{\cra c_j \cra c_i}{0_b}
    }{
    \iner{0_a}{0_b}
    } = (U_b^*(f)^{-1}V_a^T)_{ij}.
\end{align}
For the Thouless representation, these expressions may be rewritten entirely in terms of the Thouless amplitudes $Z_a = V_a^*(U_a^*)^{-1}$ and $Z_b = V_b^*(U_b^*)^{-1}$, leading to
\begin{align}
    \rho_{ij} =& (Z_b(1 + Z_a^\dagger Z_b)^{-1} Z_a^\dagger)_{ij},\nonumber \\
    \kappa_{ij} = & (Z_b(1 + Z_a^\dagger Z_b)^{-1} )_{ij},\nonumber \\
    -\bar\kappa_{ij}^* =& ((1 + Z_a^\dagger Z_b)^{-1} Z_a^\dagger)_{ij}.
\end{align}

The projection property of the generalized density matrix can be extended beyond single quasiparticle vacua. Consider the transition density matrix
\begin{align}
    R_{ k,  l} = \frac{\mat{\Psi}{\cra{C}_{ l} \an{C}_{ k}}{0_b}}{\iner{\Psi}{0_b}}, \quad k,l=1,...,2M,
\end{align}
where $\ket{\Psi} = \sum_a w_a\ket{0_a}$ is a superposition of quasiparticle vacuum states (non-orthogonal to $\ket{0_b}$). Using the linearity of the matrix elements, the density matrix can be expressed as a weighted sum of the corresponding transition density matrices,
\begin{align}
    R_{k l} =& \frac{\sum_a w_a \mat{0_a}{\cra{C}_{l} \an{C}_{k}}{0_b}}{\iner{\Psi}{0_b}} \nonumber\\
    =& \frac{\sum_a w_a\iner{0_a}{0_b}\mat{0_a}{\cra{C}_{l} \an{C}_{k}}{0_b}/\iner{0_a}{0_b} }{\iner{\Psi}{0_b}}\nonumber \\
    =&\sum_a \bar w_a  R^{a}_{kl}
\end{align}
where $\bar{\omega}_a = \omega_a \iner{0_a}{0_b}/\iner{\Psi}{0_b}$ ($\sum \bar \omega_a = 1$) and
\begin{align}
    R^a = \sum_{kl} \ket{\dot{ b}_k}(f^{-1}_a)_{kl}\bra{\dot a_l}.
\end{align}
Furthermore, one finds that the product of any two such projectors satisfies
\begin{align}
    &R^{a_1} R^{a_2} \nonumber \\
    =& \sum_{a_1, a_2} \bar\omega_{a_1} \bar\omega_{a_2} \sum_{kl,ij}\left (\ket{\dot{ b}_k}(f^{-1}_{a_1})_{kl}\bra{\dot a_{1,l}}\right)\left(\ket{\dot{ b}_i}(f^{-1}_{a_2})_{ij}\bra{\dot a_{2,j}}\right) \nonumber \\
    =&\sum_{a_1, a_2} \bar\omega_{a_1} \bar\omega_{a_2} \sum_{kl,ij} \ket{\dot{ b}_k}(f^{-1}_{a_1})_{kl}(f_{a_1})_{li}(f^{-1}_{a_2})_{ij}\bra{\dot a_{2,j}} \nonumber \\
    =&\sum_{a_1, a_2} \bar\omega_{a_1} \bar\omega_{a_2} \sum_{ij}\ket{\dot{ b}_i}(f^{-1}_{a_2})_{ij}\bra{\dot a_{2,j}} \nonumber \\
    =& R^{a_2},
\end{align}
which immediately leads to
\begin{align}
    R^2 = \sum_{a,b} \bar \omega_a \bar \omega_b R^a R^b = \sum_{a,b}\bar \omega_a \bar \omega_b R^b = R.
\end{align}
Therefore, the generalized one-body density matrix retains its projection property even when the left state (the trial wave function) is an arbitrary linear combination of quasiparticle vacua, provided that the right state (the stochastic walker) remains a quasiparticle vacuum. This result justifies the use of the generalized Wick theorem and the mean-field gauge formalism developed in the main text for correlated trial wave functions, such as symmetry-projected HFB states.

\section{Overlap calculation}
Let $\ket{\Psi_a}$ and $\ket{\Psi_b}$ be two HFB wave functions associated with quasiparticle operators $\cra B_a$ and $\cra B_b$, respectively. Throughout this appendix, we assume that both quasiparticle bases are generated by unitary Bogoliubov transformations,
\begin{align}
    \cra B_i =
    \begin{bmatrix}
        \cra c & \an c^T
    \end{bmatrix} \m W_i = \begin{bmatrix}
        \cra c & \an c^T
    \end{bmatrix}
    \begin{bmatrix}
        U_i & V_i^* \\ V_i & U_i^*
    \end{bmatrix}
\end{align}
with ${\m W}_i^\dagger = {\m W}^{-1}_i$ for $i = a,b$. As discussed in Sec. III, an HFB wave function may be represented either in the product-state form
\begin{align}
    \ket{\Psi^{P}_i} = \prod_k \an \beta_k^{(i)} \ket{0_c},
\end{align}
or in the Thouless form
\begin{align}
    \ket{\Psi^{T}_i} = \exp(\frac{1}{2}\cra c_k Z^{(i)}_{kl} \cra c_l) \ket{ 0_c},
\end{align}
where $Z^{(i)} = V_i^*(U_i^*)^{-1}$ is a skew-symmetric matrix and $\ket{0_c}$ denotes the vacuum annihilated by the operators $\an c$. 

For the product-state representation, the overlap between two HFB wave functions is given by Ref.~\cite{bertsch2012symmetry}
\begin{align}\label{olp_prod_1}
    \iner{\Psi_a^P}{\Psi_b^P} = s_M{\rm pf}\left(
    \begin{bmatrix}
        V_a^TU_a & V_a^T  V_{b}^* \\
        - V_{b}^*V_a &  U_{b}^\dagger  V_{b}^*
    \end{bmatrix}
    \right),
\end{align}
where $s_M = (-1)^{M(M-1)/2}$ and ${\rm pf}(S)$ is the Pfaffian of a skew symmetry matrix $S$. In particular, when $\ket{\Psi_a} = \ket{0_c}$, the overlap reduces to
\begin{align}\label{olp_prod_2}
    \iner{0_c}{\Psi^P_a} = {\rm pf}\left(U_{b}^\dagger  V_{b}^*\right),
\end{align}
which is used in the evaluation of normalization factors discussed in Sec. III.

For the Thouless representation, the overlap takes the well-known form~\cite{robledo2009sign}
\begin{align}
    \iner{\Psi_a^T}{\Psi_b^T} = (-1)^{M(M+1)/2}{\rm pf}\left(
    \begin{bmatrix}
        Z^b & -I \\
        I & -Z^{a*}
    \end{bmatrix}
    \right).
\end{align}
These expressions provide the overlaps required for the importance sampling, hybrid energy, and normalization factors employed throughout the HFB-AFQMC algorithm.

\section{Normalization factor calculation}
In the product-state representation, the normalization factor introduced in Sec. III is defined as
\begin{align}
    \mathcal{N} = \frac{
    \mat{\Phi}{e^{\an S}}{  \Psi_i^P}
    }{
    \iner{\Phi}{\Psi_{i+1}^P}
    },
\end{align}
where $\Phi$ is an arbitrary wave function not orthogonal to the propagated $\Psi_{i+1}^P$. The denominator is the overlap between quasiparticle vacua and can be evaluated using Eqs.~(B4--B6). For the numerator, we choose the reference vacuum $\ket{\Phi} = \ket{\Psi_0}$. In this case, the generalized one-body propagator admits the factorization with respect to the vacuum state $\ket{\Psi_0}$~\cite{balian1969nonunitary}
\begin{align}\label{c2}
    e^{\an S} = \sqrt{\det{(\m{U}_s)}}e^{\an Z}
    e^{ \an X}
    e^{ \an Y},
\end{align}
where the operators $\an X, \an Y, \an Z$ are defined by
\begin{align}
    \an Z =& \frac{1}{2} \cra{\beta}{}^{(0)} Z  (\cra \beta{}^{(0)})^T, \nonumber\\
    \an X =& \cra \beta{}^{(0)} X \an \beta^{(0)}, \nonumber \\
    \an Y =& \frac{1}{2} \an \beta^{(0)})^T Y\an \beta^{(0)}
\end{align}
with the corresponding matrices given by
\begin{align}\label{c4}
    Z = \m{V}^*_s (\m{U}_s^*)^{-1}, 
    X = -{\rm ln}((\m U_s)^\dagger),
    Y = (\m U_s^*)^{-1} \overline{\m V}_s.
\end{align}
Since both $\an Z$ and $\an X$ annihilate the left reference vacuum, only the operator $e^{\an Y}$ contributes to the matrix element, yielding
\begin{align}
    \mat{\Psi_0}{e^{\an S}}{\Psi_i} = \sqrt{\det{(\m{U}_s)}}\mat{\Psi_0}{e^{\an Y}}{\Psi_i}.
\end{align}
Note that $e^{\an Y}\ket{\Psi_0} = \ket{\Psi_0}$ and from Eq.~(\ref{tran}), The action of $e^{\an Y}$ on the quasiparticle operators give
\begin{align}
    (\an \beta^{(i)}{}')^T = e^{\an Y} (\an \beta^{(i)})^T e^{- \an Y} = \an B_0 
    \begin{bmatrix}
        V_i^*{}'\\ U_i^*{}'
    \end{bmatrix}
\end{align}
with
\begin{align}
    \begin{bmatrix}
        V_i^*{}'\\ U_i^*{}'
    \end{bmatrix}=
    \begin{bmatrix}
        I & 0 \\
        Y & I
    \end{bmatrix}
    \begin{bmatrix}
        V_i^* \\U_i^*
    \end{bmatrix}
    =\begin{bmatrix}
        V_i^* \\
        YV_i^* + U_i^*
    \end{bmatrix}.
\end{align}
Using Eqs.~(25) and (\ref{c4}), the transformed Bogoliubov matrix simplifies to
\begin{align}
    U_i^*{}' = (\m U_s^*)^{-1}(\overline{\m V}_sV_i^* + \m U_s^*U_i^*) = (\m U_s^*)^{-1}U_{i+1}^*.
\end{align}
The normalization factor can therefore be expressed entirely in terms of overlaps with the reference vacuum using Eq.~(\ref{olp_prod_2}),
\begin{align}
    \m N =& \frac{
    \mat{\Psi_0}{e^{\an S}}{  \Psi_i^P}
    }{
    \iner{\Psi_0}{\Psi_{i+1}^P}
    }\nonumber \\ 
    =& \sqrt{\det{(\m{U}_s)}}
    \frac{
    \iner{\Psi_0}{\Psi_{i}^P{}'}
    }{
    \iner{\Psi_0}{\Psi_{i+1}^P}
    } \nonumber \\
    =& 
    \sqrt{\det{(\m{U}_s)}}
    \frac{
    {\rm pf}(U_i^\dagger{}' V_i^*{}')
    }{
    {\rm pf}(U_{i+1}^\dagger V_{i+1}^*)
    }\nonumber \\
    =& 
    \sqrt{\det{(\m{U}_s)}}
    \frac{
    {\rm pf}(U_{i+1}^\dagger (\m U_s^\dagger)^{-1}V_i^*)
    }{
    {\rm pf}(U_{i+1}^\dagger V_{i+1}^*)
    }.
\end{align}
In the Thouless state representation, the normalization factor is defined by a pair of biorthogonal Fock state and one can directly employ Eq~(\ref{c2}) to factorize the exponential operator in the new transformed basis $e^{\an Z_i} \cra B_0 e^{-\an Z_i}$ and the resulting matrix element reduces immediately to the constant prefactor,
\begin{align}
    \m N = \mat{\Psi_0}{e^{\an S}}{ \Psi_i} = \sqrt{\det{(U^*)}},
\end{align}
where
\begin{align}
    U^* = \overline{\m V}_s Z_i + \m U_s^*.
\end{align}
Unlike the product-state representation, the normalization factor is therefore obtained directly from the Thouless matrix, without requiring the evaluation of Pfaffian overlaps. This compact expression is one of the practical advantages of the Thouless representation.

\section{Ito-stochastic differential equations for the pairing system}
By discarding irrelevant constant terms, the Richardson Hamiltonian can be rewritten in the general quadratic form introduced in Eq.~(\ref{ham_qmc}),
\begin{align}
     \an H = \an{\mathcal{T}} - \frac{1}{2}  \sum_{a = x,y}\lambda_a \an{\mathcal{O}}_a^2,
\end{align}
where $\lambda_x = -\lambda_y = G/2$ and
\begin{align}
    \an{\m T} =& \frac{1}{2}\sum_{p,\sigma=\uparrow\downarrow}\widetilde{\varepsilon}_p( \cra c_{p\sigma}\an c_{p \sigma} - \an c_{p \sigma}\cra c_{p\sigma}),\nonumber \\
    \an{\m O}_x =&\sum_p \cra c_{p\uparrow} \cra c_{p\downarrow} + \an c_{p\downarrow} \an c_{p \uparrow},\nonumber \\
    \an{\m O}_y =& \sum_p \cra c_{p\uparrow} \cra c_{p\downarrow} - \an c_{p\downarrow} \an c_{p \uparrow},
\end{align}
with $\widetilde{\varepsilon}_p = \varepsilon_p - \mu - G/2$, where $\mu$ denotes the chemical potential.

For this model, the stochastic walkers remain within the BCS manifold throughout the imaginary-time evolution and can be parametrized by the Bogoliubov amplitudes $u_i$ and $v_i$ as
\begin{align}
    \ket{\Psi} = \sum_{i = 1}^M (u_i + v_i \cra c_{i\uparrow} \cra c_{i\downarrow})\ket{0_c}.
\end{align}
The stochastic evolution of these amplitudes can be derived explicitly using Ito calculus for the different stochastic propagators introduced in Sec. III C. The resulting equations provide a transparent connection between the stochastic gauge formulation and the corresponding Brownian motion of quasiparticle states.

We first consider the stochastic propagator incorporating the drift gauge introduced in Eq. (38),
\begin{align}
    e^{d\an S} = \exp\left(
    -d\tau \an{\m T} - d\tau \lambda_a g_a \an{\m O}_a + dW_a \sqrt{\lambda_a} \an{\m O}_a
    \right),
\end{align}
where $dW_a$ denotes independent Wiener increments satisfying Ito's rule $dW_i dW_j = d\tau \delta_{ij}$. Using the algebra of quasiparticle operators together with Ito calculus~\cite{gardiner1985handbook}, one obtains the stochastic differential equations governing the BCS amplitudes,
\begin{widetext}
\begin{align}
    d\begin{bmatrix}
        v_i \\ u_i
    \end{bmatrix}
    =
    -d\tau 
    \begin{bmatrix}
        \tilde{\varepsilon}_i & - \frac{G}{2}(g_x + g_y) \\
        -\frac{G}{2}(g_x - g_y) & -\tilde{\varepsilon}_i
    \end{bmatrix}
    \begin{bmatrix}
        v_i \\ u_i
    \end{bmatrix}
    +\sum_{a = x,y}dW_a \sqrt{\lambda_a} \m{O}_a
    \begin{bmatrix}
        v_i \\ u_i
    \end{bmatrix}, \quad
    i = 1,...,M,
\end{align}
\end{widetext}
where
\begin{align}
    \m O_x = \begin{bmatrix}
        0 & 1 \\
        1 & 0
    \end{bmatrix},
    \qquad
    \m O_y = \begin{bmatrix}
        0 & 1 \\
        -1 & 0
    \end{bmatrix}.
\end{align}
The corresponding evolution of the walker weight is given by
\begin{align}
    d\Omega/\Omega = -d\tau \av{\an H}_{T,s} + dW_a \sqrt{\lambda_a}(\av{\an{\m O}_a}_{T,s} + g_a).
\end{align}
A particularly important choice is the conventional force bias, $g_a = - \av{\an{\m O}_a}_{T,s}$, which minimizes the diffusion of the walker weights. Under this choice, the deterministic drift matrix $A_d$ reduces to the standard BCS mean-field Hamiltonian,
\begin{align}
    A_d = \begin{bmatrix}
        \varepsilon_i - \tilde{\mu} & \Delta(\kappa) \\
        \Delta(\bar \kappa^*)& - (\varepsilon_i - \tilde{\mu})
    \end{bmatrix},
\end{align}
with the BCS chemical potential $\tilde{\mu} = \mu + G/2$ and the pairing field
\begin{align}
    \Delta(\kappa) = G\sum_i \kappa_i =  G\sum_i \av{\cra c_{i\uparrow} \cra c_{i\downarrow}}_{T,s},\nonumber \\
    \Delta(\bar \kappa^*) = G\sum_i \bar \kappa^* = G\sum_i \av{\an c_{i \downarrow} \an c_{i \uparrow}}_{T,s} 
\end{align}
Thus, the conventional phaseless AFQMC can be interpreted as a Brownian motion of quasiparticle states evolving under the standard BCS mean-field Hamiltonian.

An alternative choice is to minimize the stochastic fluctuations of the walker trajectories instead of the weight diffusion. This is achieved by minimizing
$||\big(\an{\m O}_a + g_a\big)\ket{\Psi}||^2
= |g_a + \av{\an{\m O}_a}_{s,s}|^2 + \av{\an{\m O}_a^\dagger \an{\m O}_a}_{s,s} -\av{\an{\m O}_a^\dagger}_{s,s} \av{\an{\m O}_a}_{s,s}$
which yields $g_a = -\langle \hat{\m O}_a \rangle_{s,s}$, where $\av{\cdot}_{s,s}$ denotes the expectation value evaluated with respect to the walker itself. The resulting drift matrix has the same form as Eq. (D9), except that the pairing field is constructed from the walker density matrix rather than the mixed density matrix, namely $\kappa_i = \av{\cra c_{i\uparrow} \cra c_{i \downarrow}}_{s,s}$ and $\bar \kappa^* = \kappa^*$.

The MF-AFQMC formulation retains the same local-energy expression for the weight evolution as the conventional phaseless AFQMC but modifies the walker propagation by replacing the conventional drift with the mean-field gauge introduced in Sec. III C. The corresponding stochastic differential equations for the BCS amplitudes become
\begin{widetext}
    \begin{align}
    d
    \begin{bmatrix}
        v_i \\ u_i
    \end{bmatrix}
    = -d\tau 
    \begin{bmatrix}
        \tilde{\varepsilon}_i - G(\rho_i - 1/2) & \Delta(\kappa)\\
        \Delta(\bar \kappa)^* & -\tilde{\varepsilon}_i + G(\rho_i - 1/2)
    \end{bmatrix}
    \begin{bmatrix}
        v_i \\ u_i
    \end{bmatrix}
    + \sum_{a=x,y}dW_a\, \sqrt{\lambda_a} (I - R) O_a
    \begin{bmatrix}
        v_i \\ u_i
    \end{bmatrix}, \quad i = 1,...,  M
\end{align}
\end{widetext}
where $R_i$ denotes the generalized one-body density matrix,
\begin{align}
    R_i = \begin{bmatrix}
        \rho_i & \kappa_i\\
        \bar \kappa^*_i & 1- \rho_i
    \end{bmatrix},
\end{align}
with $\rho_i = \av{\cra c_{i\sigma} \an c_{i\sigma} }_{T,s}$, $\kappa_i= \av{\an c_{i \downarrow} \an c_{i\uparrow}}_{T,s}$, and $\bar \kappa^*_i = \av{\cra c_{i \uparrow} \cra c_{i \downarrow}}_{T,s}$. Compared with the conventional formulation, the mean-field gauge incorporates the Hartree-Fock contribution, $-G(\rho_i - 1/2)$, into the diagonal components of the deterministic drift matrix while restricting the stochastic diffusion to the residual particle-hole channels through the projector $(I - R)$. For the Richardson model considered in the present work, this formulation provides an alternative interpretation of the stochastic evolution in the BCS manifold, although, as discussed in Sec. IV, it leads to numerical results that are comparable to those obtained with the conventional force-bias formulation.

\bibliography{bibliography}

@PREAMBLE{
 "\providecommand{\noopsort}[1]{}" 
 # "\providecommand{\singleletter}[1]{#1}%" 
}

@article{mywork,
  title={Gauge Auxiliary-Field Quantum {M}onte {C}arlo Method for Many-Fermion Systems},
  author={Zhang, Zhaozhan},
  journal={arXiv preprint arXiv:2607.20917},
  year={2026}
}

@article{hu2026stochastic,
  title={Stochastic Similarity Renormalization Group},
  author={Hu, Rongzhe and Zhen, Xin and Xu, Furong and Pei, Junchen},
  journal={arXiv preprint arXiv:2607.08830},
  year={2026}
}

@article{hu2025ab,
  title={Ab initio exact calculation of strongly-correlated nucleonic matter},
  author={Hu, R and Jin, S and Zhen, X and Shang, H and Pei, J and Xu, F and Marino, F},
  journal={arXiv preprint arXiv:2508.09252},
  year={2025}
}

@article{sukurma2024toward,
  title={Toward large-scale afqmc calculations: Large time step auxiliary-field quantum monte carlo},
  author={Sukurma, Zoran and Schlipf, Martin and Humer, Moritz and Taheridehkordi, Amir and Kresse, Georg},
  journal={Journal of Chemical Theory and Computation},
  volume={20},
  number={10},
  pages={4205--4217},
  year={2024},
  publisher={ACS Publications}
}

@article{bertsch2012symmetry,
  title={Symmetry restoration in {H}artree-{F}ock-{B}ogoliubov based theories},
  author={Bertsch, George F and Robledo, Luis M},
  journal={Physical Review Letters},
  volume={108},
  number={4},
  pages={042505},
  year={2012},
  publisher={APS}
}

@article{robledo2009sign,
  title={Sign of the overlap of {H}artree-{F}ock-{B}ogoliubov wave functions},
  author={Robledo, Luis M},
  journal={Physical Review C—Nuclear Physics},
  volume={79},
  number={2},
  pages={021302},
  year={2009},
  publisher={APS}
}

@article{lee2020utilizing,
  title={Utilizing essential symmetry breaking in auxiliary-field quantum {M}onte {C}arlo: Application to the spin gaps of the {C}$_{36}$ fullerene and an iron porphyrin model complex},
  author={Lee, Joonho and Malone, Fionn D and Morales, Miguel A},
  journal={Journal of Chemical Theory and Computation},
  volume={16},
  number={5},
  pages={3019--3027},
  year={2020},
  publisher={ACS Publications}
}

@article{xu2024coexistence,
  title={Coexistence of superconductivity with partially filled stripes in the {H}ubbard model},
  author={Xu, Hao and Chung, Chia-Min and Qin, Mingpu and Schollw{\"o}ck, Ulrich and White, Steven R and Zhang, Shiwei},
  journal={Science},
  volume={384},
  number={6696},
  pages={eadh7691},
  year={2024},
  publisher={American Association for the Advancement of Science}
}

@article{shi2017many,
  title={Many-body computations by stochastic sampling in {H}artree-{F}ock-{B}ogoliubov space},
  author={Shi, Hao and Zhang, Shiwei},
  journal={Physical Review B},
  volume={95},
  number={4},
  pages={045144},
  year={2017},
  publisher={APS}
}

@article{vitali2024monte,
  title={{M}onte {C}arlo methods in the manifold of {H}artree-{F}ock-{B}ogoliubov wave functions},
  author={Vitali, Ettore and Rosenberg, Peter and Zhang, Shiwei},
  journal={The Journal of Chemical Physics},
  volume={161},
  number={13},
  year={2024},
  publisher={AIP Publishing}
}

@article{juillet2017phaseless,
  title={Phaseless quantum {M}onte-{C}arlo approach to strongly correlated superconductors with stochastic {H}artree-{F}ock-{B}ogoliubov wavefunctions},
  author={Juillet, Olivier and Lepr{\'e}vost, Alexandre and Bonnard, J{\'e}r{\'e}my and Fr{\'e}sard, Raymond},
  journal={Journal of Physics A: Mathematical and Theoretical},
  volume={50},
  number={17},
  pages={175001},
  year={2017},
  publisher={IOP Publishing}
}

@article{balian1969nonunitary,
  title={Nonunitary {B}ogoliubov transformations and extension of {W}ick’s theorem},
  author={Balian, R and Brezin, E},
  journal={Il Nuovo Cimento B (1965-1970)},
  volume={64},
  number={1},
  pages={37--55},
  year={1969},
  publisher={Springer}
}

@article{PhysRev.101.1730,
  title = {Natural Orbitals in the Quantum Theory of Two-Electron Systems},
  author = {L\"owdin, Per-Olov and Shull, Harrison},
  journal = {Physical Review},
  volume = {101},
  issue = {6},
  pages = {1730--1739},
  numpages = {0},
  year = {1956},
  month = {Mar},
  publisher = {American Physical Society}
}

@article{PhysRevLett.111.012502,
  title = {Constrained-Path Quantum {M}onte {C}arlo Approach for the Nuclear Shell Model},
  author = {Bonnard, J. and Juillet, O.},
  journal = {Physical Review Letters},
  volume = {111},
  issue = {1},
  pages = {012502},
  numpages = {5},
  year = {2013},
  month = {Jul},
  publisher = {American Physical Society},
  doi = {10.1103/PhysRevLett.111.012502},
  url = {https://link.aps.org/doi/10.1103/PhysRevLett.111.012502}
}

@article{zhang2003quantum,
  title={Quantum {M}onte {C}arlo method using phase-free random walks with {S}later determinants},
  author={Zhang, Shiwei and Krakauer, Henry},
  journal={Physical Review Letters},
  volume={90},
  number={13},
  pages={136401},
  year={2003},
  publisher={APS}
}

@article{corney2006gaussian,
  title={Gaussian phase-space representations for fermions},
  author={Corney, Joel F and Drummond, Peter D},
  journal={Physical Review B},
  volume={73},
  number={12},
  pages={125112},
  year={2006},
  publisher={APS}
}

@article{corney2005gaussian,
  title={Gaussian operator bases for correlated fermions},
  author={Corney, JF and Drummond, PD},
  journal={Journal of Physics A: Mathematical and General},
  volume={39},
  number={2},
  pages={269},
  year={2005},
  publisher={IOP Publishing}
}

@article{lee2022twenty,
  title={Twenty years of auxiliary-field quantum {M}onte {C}arlo in quantum chemistry: An overview and assessment on main group chemistry and bond-breaking},
  author={Lee, Joonho and Pham, Hung Q and Reichman, David R},
  journal={Journal of Chemical Theory and Computation},
  volume={18},
  number={12},
  pages={7024--7042},
  year={2022},
  publisher={ACS Publications}
}

@article{shi2021some,
  title={Some recent developments in auxiliary-field quantum {M}onte {C}arlo for real materials},
  author={Shi, Hao and Zhang, Shiwei},
  journal={The Journal of Chemical Physics},
  volume={154},
  number={2},
  year={2021},
  publisher={AIP Publishing}
}

@article{motta2018ab,
  title={Ab initio computations of molecular systems by the auxiliary-field quantum {M}onte {C}arlo method},
  author={Motta, Mario and Zhang, Shiwei},
  journal={Wiley Interdisciplinary Reviews: Computational Molecular Science},
  volume={8},
  number={5},
  pages={e1364},
  year={2018},
  publisher={Wiley Online Library}
}

@article{zhang1997constrained,
  title={Constrained path {M}onte {C}arlo method for fermion ground states},
  author={Zhang, Shiwei and Carlson, Joseph and Gubernatis, James E},
  journal={Physical Review B},
  volume={55},
  number={12},
  pages={7464},
  year={1997},
  publisher={APS}
}

@article{hubbard1959calculation,
  title={Calculation of partition functions},
  author={Hubbard, John},
  journal={Physical Review Letters},
  volume={3},
  number={2},
  pages={77},
  year={1959},
  publisher={APS}
}

@article{thouless1960stability,
  title={Stability conditions and nuclear rotations in the {H}artree-{F}ock theory},
  author={Thouless, David J},
  journal={Nuclear Physics},
  volume={21},
  pages={225--232},
  year={1960},
  publisher={Elsevier}
}

@article{thouless1961vibrational,
  title={Vibrational states of nuclei in the random phase approximation},
  author={Thouless, DJ},
  journal={Nuclear Physics},
  volume={22},
  number={1},
  pages={78--95},
  year={1961},
  publisher={Elsevier}
}

@article{trotter1959product,
  title={On the product of semi-groups of operators},
  author={Trotter, Hale F},
  journal={Proceedings of the American Mathematical Society},
  volume={10},
  number={4},
  pages={545--551},
  year={1959},
  publisher={JSTOR}
}

@book{gardiner1985handbook,
  title={Handbook of stochastic methods},
  author={Gardiner, Crispin W and others},
  volume={3},
  year={1985},
  publisher={Springer Berlin}
}

@inproceedings{bonnard2016constrained,
  title={A constrained-path quantum {M}onte-{C}arlo approach for the nuclear shell model},
  author={Bonnard, J{\'e}r{\'e}my and Juillet, Olivier},
  booktitle={Journal of Physics: Conference Series},
  volume={724},
  pages={012004},
  year={2016},
  organization={IOP Publishing}
}

@article{bohr1958possible,
  title={Possible analogy between the excitation spectra of nuclei and those of the superconducting metallic state},
  author={Bohr, Aage and Mottelson, Ben R and Pines, David},
  journal={Physical Review},
  volume={110},
  number={4},
  pages={936},
  year={1958},
  publisher={APS}
}

@article{richardson1964exact,
  title={Exact eigenstates of the pairing-force {H}amiltonian},
  author={Richardson, RW and Sherman, Noah},
  journal={Nuclear Physics},
  volume={52},
  pages={221--238},
  year={1964},
  publisher={Elsevier}
}

@article{richardson1965number,
  title={Number dependence of the accuracy of {BCS} calculations in the theory of deformed nuclei},
  author={Richardson, RW},
  journal={Physics Letters},
  volume={14},
  number={4},
  pages={325--327},
  year={1965},
  publisher={Elsevier}
}

@inproceedings{de2014probing,
  title={Probing pairing correlations in {S}n isotopes using {R}ichardson-{G}audin integrability},
  author={De Baerdemacker, Stijn and Hellemans, Veerle and van den Berg, Rianne and Caux, Jean-S{\'e}bastien and Heyde, Kristiaan and Van Raemdonck, Mario and Van Neck, Dimitri and Johnson, Paul A},
  booktitle={Journal of Physics: Conference Series},
  volume={533},
  pages={012058},
  year={2014},
  organization={IOP Publishing}
}

@article{richardson1963application,
  title={Application to the exact theory of the pairing model to some even isotopes of lead},
  author={Richardson, RW},
  journal={Physics Letter},
  volume={5},
  number={1},
  pages={82-84},
  year={1963},
  publisher={Elsevier}
}

@article{companys2024eigenvector,
  title={Eigenvector continuation for the pairing Hamiltonian},
  author={Companys Franzke, Margarida and Tichai, Alexander and Hebeler, Kai and Schwenk, Achim},
  journal={Physical Review C},
  volume={109},
  number={2},
  pages={024311},
  year={2024},
  publisher={APS}
}

@article{rigo2023solving,
  title={Solving the nuclear pairing model with neural network quantum states},
  author={Rigo, Mauro and Hall, Benjamin and Hjorth-Jensen, Morten and Lovato, Alessandro and Pederiva, Francesco},
  journal={Physical Review E},
  volume={107},
  number={2},
  pages={025310},
  year={2023},
  publisher={APS}
}

@article{brolli2025diagrammatic,
  title={Diagrammatic {M}onte {C}arlo for finite systems at zero temperature},
  author={Brolli, Stefano and Barbieri, Carlo and Vigezzi, Enrico},
  journal={Physical Review Letters},
  volume={134},
  number={18},
  pages={182502},
  year={2025},
  publisher={APS}
}

@book{TB:ring2004nuclear,
  title={The nuclear many-body problem},
  author={Ring, Peter and Schuck, Peter},
  year={2004},
  publisher={Springer Science \& Business Media}
}

@article{lee2020performance,
  title={The performance of phaseless auxiliary-field quantum {M}onte {C}arlo on the ground state electronic energy of benzene},
  author={Lee, Joonho and Malone, Fionn D and Reichman, David R},
  journal={The Journal of Chemical Physics},
  volume={153},
  number={12},
  year={2020},
  publisher={AIP Publishing}
}

\end{document}